\documentclass[twocolumn]{aastex63}
    
\usepackage{amssymb,amsmath,amsthm,amstext}
\usepackage{graphicx}
\usepackage{mathrsfs}
\usepackage{booktabs}
\usepackage{multirow}
\usepackage{empheq}
\usepackage{verbatim}
\usepackage{mathtools}
\usepackage{hyperref}
\usepackage{xspace}
\usepackage{xcolor}

\defcitealias{Pesce_2018}{P18}
\defcitealias{Gao_2017}{G17}

\def\j0437{J0437+2456\xspace}       % J0437 shorthand
\def\HI{H\textsc{i}\xspace}             % HI (neutral hydrogen)
\def\kms{km\,s$^{-1}$\xspace}       % km/s
\def\h2o{H$_2$O\xspace}       % km/s
\def\dynesty{\texttt{dynesty}\xspace}       % dynesty shorthand
\def\atlas{ATLAS$^{\text{3D}}$\xspace}       % ATLAS3D shorthand

\begin{document}

\title{A restless supermassive black hole in the galaxy J0437+2456}
\shorttitle{A restless supermassive black hole in the galaxy J0437+2456}

\correspondingauthor{Dominic~W.~Pesce}
\email{dpesce@cfa.harvard.edu}

\author[0000-0002-5278-9221]{Dominic~W.~Pesce}
\affiliation{Center for Astrophysics $|$ Harvard \& Smithsonian, 60 Garden Street, Cambridge, MA 02138, USA}
\affiliation{Black Hole Initiative at Harvard University, 20 Garden Street, Cambridge, MA 02138, USA}

\author[0000-0003-0248-5470]{Anil~C.~Seth}
\affiliation{Department of Physics and Astronomy, University of Utah, 115 South 1400 East, Salt Lake City, UT 84112, USA}

\author{Jenny~E.~Greene}
\affiliation{Department of Astrophysics, Princeton University, Princeton, NJ, USA}

\author[0000-0002-1468-9203]{James~A.~Braatz}
\affiliation{National Radio Astronomy Observatory, 520 Edgemont Road, Charlottesville, VA 22903, USA}

\author[0000-0003-4724-1939]{James~J.~Condon}
\affiliation{National Radio Astronomy Observatory, 520 Edgemont Road, Charlottesville, VA 22903, USA}

\author[0000-0002-8990-1811]{Brian~R.~Kent}
\affiliation{National Radio Astronomy Observatory, 520 Edgemont Road, Charlottesville, VA 22903, USA}

\author[0000-0002-0470-6540]{Davor Krajnovi{\'c}}
\affiliation{Leibniz-Institut f\"ur Astrophysik Potsdam (AIP), An der Sternwarte 16, D-14482 Potsdam, Germany}

\begin{abstract}
We present the results from an observing campaign to confirm the peculiar motion of the supermassive black hole (SMBH) in \j0437 first reported in \cite{Pesce_2018}.  Deep observations with the Arecibo Observatory have yielded a detection of neutral hydrogen (\HI) emission, from which we measure a recession velocity of 4910\,\kms for the galaxy as a whole.  We have also obtained near-infrared integral field spectroscopic observations of the galactic nucleus with the Gemini North telescope, yielding spatially resolved stellar and gas kinematics with a central velocity at the innermost radii ($0.1^{\prime \prime} \approx 34$\,pc) of 4860\,\kms.  Both measurements differ significantly from the $\sim$4810\,\kms H$_2$O megamaser velocity of the SMBH, supporting the prior indications of a velocity offset between the SMBH and its host galaxy.  However, the two measurements also differ significantly from one another, and the galaxy as a whole exhibits a complex velocity structure that implies the system has recently been dynamically disturbed.  These results make it clear that the SMBH is not at rest with respect to the systemic velocity of the galaxy, though the specific nature of the mobile SMBH -- i.e., whether it traces an ongoing galaxy merger, a binary black hole system, or a gravitational wave recoil event -- remains unclear.
\end{abstract}

\section{Introduction}

Given that nearly all galaxies are thought to harbor central supermassive black holes \citep[SMBHs;][]{Magorrian_1998}, interactions between SMBHs have long been recognized as a natural and perhaps inevitable byproduct of galaxy mergers.  The two primary dynamical states that result from such interactions are SMBH binaries \citep{Begelman_1980,Roos_1981} and gravitational recoil events \citep{Fitchett_1983,Redmount_1989}, both of which predict substantial nonequilibrium (``peculiar'') motion of the SMBH with respect to its surrounding environment.  Yet despite much theoretical attention and observational effort, direct dynamical evidence for SMBH peculiar motion has remained elusive \citep[see, e.g.,][]{Eracleous_2012,Popovic_2012,Komossa_2016,Barack_2019}.  In the absence of recent interactions with comparable-mass objects, an SMBH is expected to be in kinetic equilibrium with its surrounding environment \citep{Merritt_2007}; for most SMBHs, the equilibrium velocity is $\ll 1$\,\kms with respect to the system barycenter.

\citet[][hereafter \citetalias{Pesce_2018}]{Pesce_2018} presented a technique for using H$_2$O megamasers to measure SMBH peculiar motions.  The key idea is that masers residing in the accretion disks around SMBHs (on scales of $\sim$0.1\,pc) act as test particles whose dynamics can be used to probe the gravitational potential around the black hole, and very long baseline interferometric (VLBI) maps of the maser distribution enable precise (uncertainty ${\lesssim}10$\,\kms) measurements of the SMBH's line-of-sight velocity \citep[e.g.,][]{Miyoshi_1995,Kuo_2011,Gao_2017}.  \citetalias{Pesce_2018} compared the maser-derived SMBH velocity measurements for 10 systems with independent estimates of their host galaxy velocities to constrain relative motions.  One galaxy from the \citetalias{Pesce_2018} sample -- SDSS J043703.67+245606.8, hereafter \j0437 -- showed a statistically significant (${>}5{\sigma}$) difference between the SMBH and host galaxy line-of-sight velocities; \citetalias{Pesce_2018} thus identified \j0437 as a promising candidate for hosting either a recoiling or binary SMBH.

\j0437 is an approximately Sb-type spiral galaxy located at a distance of ${\sim}70$\,Mpc \citep{Greene_2016,Pjanka_2017}.  As measured by the Sloan Digital Sky Survey (SDSS)\footnote{Here we quote the quantities compiled in the NASA-Sloan Atlas, \url{http://nsatlas.org/}.}, \j0437 has an $r$-band absolute AB magnitude of $M_r = -21.37$ and an estimated stellar mass of $7.2 \times 10^{10}$\,M$_{\odot}$.  The megamaser system in \j0437 was mapped by \citet[][hereafter \citetalias{Gao_2017}]{Gao_2017}, who also modeled the maser rotation curve and determined an SMBH velocity of $4818 \pm 10.5$\,\kms.  \citetalias{Pesce_2018} used a SDSS spectrum to measure the recession velocity of \j0437 to be $4887.6 \pm 7.1$\,\kms.  The apparent $69.6 \pm 12.7$\,\kms blueshift of the SMBH with respect to its host galaxy constitutes the putative peculiar motion.  However, given the strong prior expectation for zero peculiar motion and the possibility that systematic effects such as SDSS fiber misalignment could plausibly account for a large fraction of the observed velocity difference, \citetalias{Pesce_2018} cautioned that the peculiar motion measurement should be regarded as tentative pending corroborating observations.

In this paper we present the results from a followup observing campaign to confirm the peculiar motion of the SMBH in \j0437.  This paper is organized as follows.  In \autoref{sec:Observations} we describe our observations and subsequent data reduction procedures, and in \autoref{sec:Analysis} we detail the velocity measurements made using these data.  We discuss the results in \autoref{sec:Discussion}, and we summarize and conclude in \autoref{sec:Conclusion}.  Unless noted otherwise, all velocities quoted in this paper use the optical convention in the barycentric reference frame, and we assume a distance to \j0437 of 70\,Mpc.

\section{Observations and data reduction} \label{sec:Observations}

\begin{figure}[t]
	\centering
		\includegraphics[width=1.00\columnwidth]{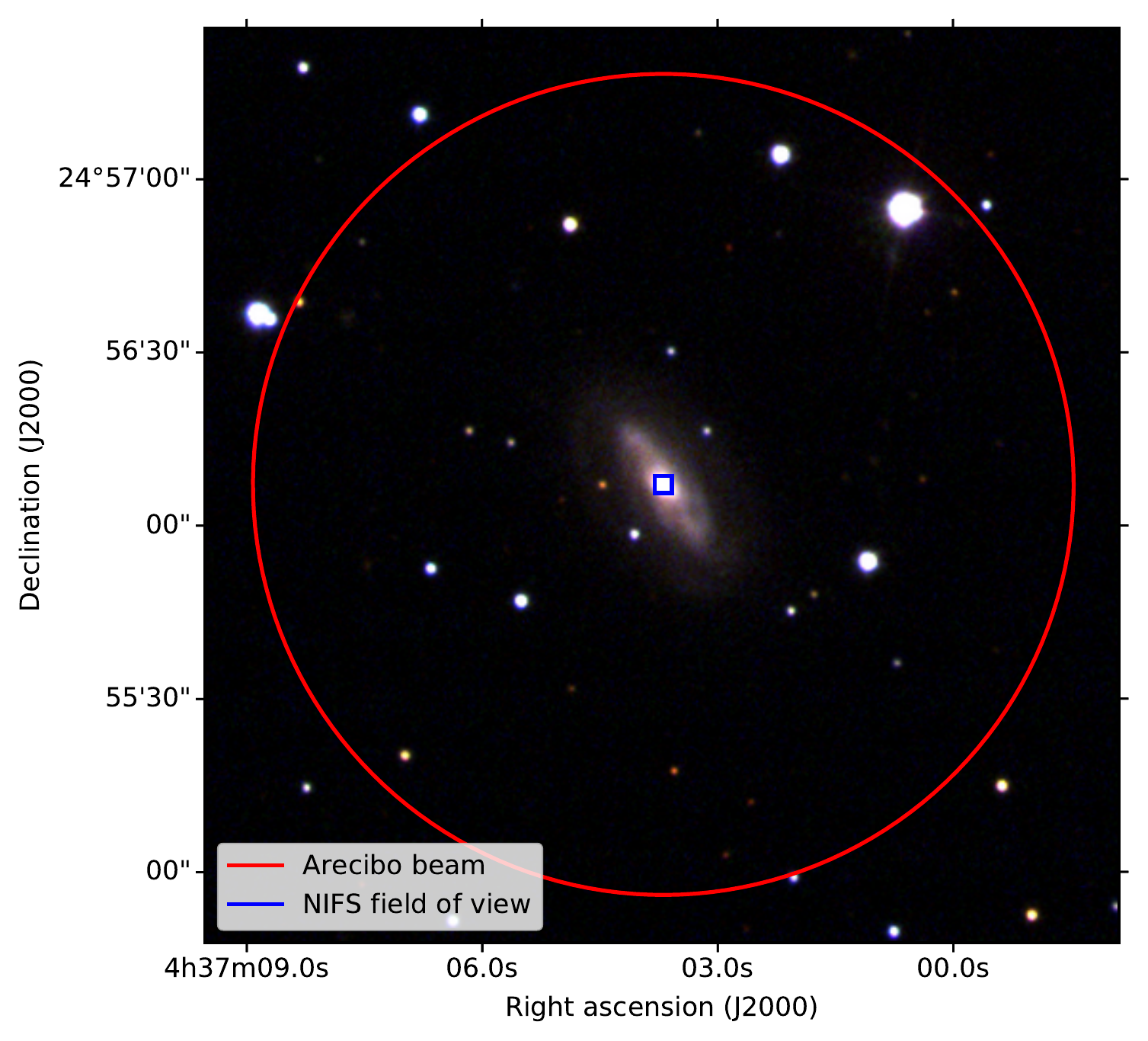}
	\caption{False-color image of \j0437 made by combining the $i$-, $r$-, and $g$-band observations from the SDSS Legacy Survey \citep{York_2000}, with the 2.4-arcminute Arecibo beam and the $3^{\prime \prime} \times 3^{\prime \prime}$ NIFS field of view overplotted in red and blue, respectively.}
	\label{fig:SDSS_image}
\end{figure}

In quiescent systems, \HI provides an appealing recession velocity tracer because it follows the global dynamics of the galaxy well outside of the SMBH sphere of influence and does not suffer from reddening or extinction.  \citetalias{Pesce_2018} targeted \j0437 with the Very Large Array (VLA) to observe neutral hydrogen (\HI), but no emission was detected within the six-hour integration time.  We have obtained followup \HI observations of \j0437 using the Arecibo Observatory, which is much more sensitive than the VLA to low surface brightness emission but which lacks the ability to spatially resolve the gas distribution (see \autoref{fig:SDSS_image}).  Our Arecibo observations are presented in \autoref{sec:Arecibo}.

Lacking \HI data, \citetalias{Pesce_2018} measured the recession velocity for \j0437 using an SDSS spectrum.  Like the Arecibo spectrum, the kinematics contributing to the SDSS spectrum are spatially unresolved within the 3-arcsecond aperture of the optical fiber used to transport light from the focal plane to the spectrograph \citep{Gunn_2006}.  However, because the aperture is smaller than the region containing the emitting material, the SDSS measurement is subject to an unknown amount of systematic uncertainty associated with the relative placement of the fiber center and the galactic nucleus.  We have thus obtained followup high-resolution integral field spectra taken using the Gemini North telescope, which are able to spatially resolve the nuclear kinematics.  Additionally, dust absorption should be weaker in the NIFS near-infrared waveband than at the SDSS optical wavelengths, so any velocity errors caused by patchy dust absorption will be smaller.  Our Gemini observations are presented in \autoref{sec:Gemini}.

\subsection{Arecibo data} \label{sec:Arecibo}

We performed \HI spectral-line observations of \j0437 over 6 nights using the Arecibo Observatory L-wide receiver.  The observations were position-switched, with 5 minutes on and 5 minutes off source at matched elevation.  We used the Wideband Arecibo Pulsar Processor (WAPP) spectrometer backend in single-polarization, 9-level autocorrelation mode with 4096 channels spanning the bandwidth 1384.5–-1409.5\,MHz (i.e., $\pm$2500\,\kms centered on the \HI line). We used two such boards, one per polarization.  Calibration diodes were observed at the end of every scan to determine the flux density scale.

\autoref{tab:AreciboObservations} lists the on-source integration times for each of the 6  nights.  With a declination of $+25$ degrees, \j0437 passes through the Arecibo observing window for ${\sim}2$ hours at a time. The position-switched observations thus yielded roughly an hour of on-source time per night.

We reduced the Arecibo data using AO IDL\footnote{\url{http://outreach.naic.edu/ao/scientist-user-portal/astronomy/IDL-Routines/Download-AO-IDL}}.  We first converted the flux scale from K to Jy using the regular gain curve monitoring scans\footnote{\url{http://www.naic.edu/\~phil/sysperf/sysperfbymon.html}} performed by the observatory (see \autoref{tab:AreciboObservations}).   Each spectral scan was Hanning smoothed to mitigate ringing, and a fourth-order polynomial fit to the emission-free regions of the spectrum was subtracted off to remove low-frequency baseline ripples.  We then combined each scan and both polarizations using an RMS-weighted average.

\begin{figure}[t]
	\centering
		\includegraphics[width=1.00\columnwidth]{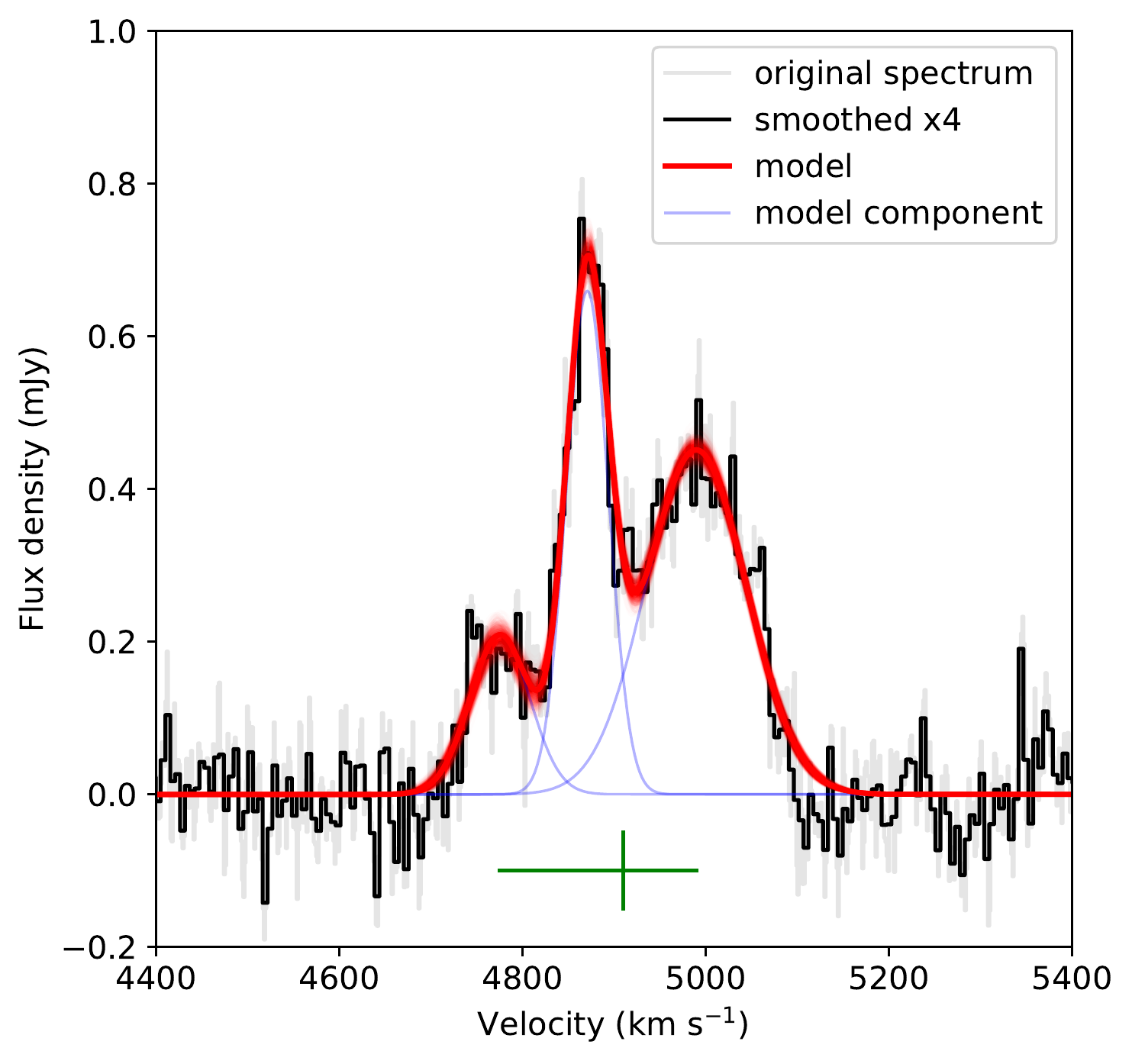}
	\caption{1.4 GHz Arecibo spectrum towards \j0437.  The spectrum is plotted with its native spectral resolution in gray, and the spectrum after smoothing by a 4-channel boxcar is shown in black.  1000 random posterior samples from a 3-component Gaussian model fit are overplotted in red, with the individual components for the best-fit model shown in blue.  The measured $V_{20}$ velocity (see \autoref{sec:HImodeling}) is marked by a vertical green line, and the peak-to-peak velocity range is marked by a horizontal green line.}
	\label{fig:HI_spectrum}
\end{figure}

\autoref{fig:HI_spectrum} shows the Arecibo spectrum, in which we strongly detect \HI emission around the expected recession velocity range.  The spectrum peaks at ${\sim}0.7$\,mJy and has an integrated flux of ${\sim}$0.12\,Jy\,\kms, consistent with the non-detection reported in \citetalias{Pesce_2018}.  Assuming the \HI is optically thin, the total \HI mass is given by \citep{Haynes_2011,Condon_2016}

\begin{equation}
M_{\text{\HI}} = \left( 2.356 \times 10^5 \text{ M}_{\odot} \right) \left( \frac{D}{\text{Mpc}} \right)^2 \left( \frac{\int S_{\nu}(v) dv}{\text{Jy\,\kms}} \right) ,
\end{equation}

\noindent where $D$ is the distance to the galaxy and $S_{\nu}(v)$ is the flux density as a function of velocity $v$.  For a distance of 70\,Mpc to \j0437, we estimate $M_{\text{\HI}} \approx 1.4 \times 10^8$\,M$_{\odot}$ (see also \autoref{sec:Analysis}).

\begin{deluxetable}{lccc}
\tablecolumns{4}
\tablewidth{0pt}
\tablecaption{Arecibo observation details\label{tab:AreciboObservations}}
\tablehead{ & \colhead{\textbf{Integration}} & \colhead{$\boldsymbol{T}_{\textbf{sys}}$} & \colhead{\textbf{Gain}} \\
\colhead{\textbf{Date}} & \colhead{(min.)} & \colhead{(K)} & \colhead{(K\,Jy$^{-1}$)}}
\startdata
2019 Jan 22 & 60 & 26.6 & 6.9 \\
2019 Jan 23 & 65 & 26.8 & 6.9 \\
2019 Jan 24 & 65 & 26.9 & 6.8 \\
2019 Feb 11 & 60 & 27.1 & 7.0 \\
2019 Feb 14 & 50 & 27.4 & 6.9 \\
2019 Feb 17 & 50 & 27.0 & 6.9
\enddata
\tablecomments{Observing dates, on-source integration times, system temperatures, and gains for the Arecibo observations.}
\end{deluxetable}

\subsection{Gemini data} \label{sec:Gemini}

We obtained integral field spectra of a $3^{\prime \prime} \times 3^{\prime \prime}$ region centered on the nucleus of \j0437 using the Gemini North Near-Infrared Integral Field Spectrometer (NIFS) on 2018 November 21 in natural seeing mode.  The spectrometer grating was set for K-band, with a central wavelength of 2.18\,$\mu$m and spanning the range 1.99--2.41\,$\mu$m.  Nine 500-second exposures were taken, with five dithered exposures on-source and four offset to a blank sky location for subtraction.  The observations were performed at airmasses of 1.2--1.6 and seeing conditions corresponding to a zenith-corrected point spread function of $\sim$0.3 arcseconds FWHM in $K$-band.

The NIFS data were reduced using the Gemini version 1.13 IRAF packages, with slight modifications to enable error array propagation and cube combination as described in \cite{Ahn_2018}.  The resulting final data cube has a central signal-to-noise of $\sim$28, dropping to $\sim$10 at 0$\farcs$5 radius.  Both strong stellar aborption lines and excited H$_2$ emission lines are seen, and their velocities are described in more detail in \autoref{sec:NIFSVelocities}.

\begin{figure*}[t]
	\centering
		\includegraphics[width=1.00\textwidth]{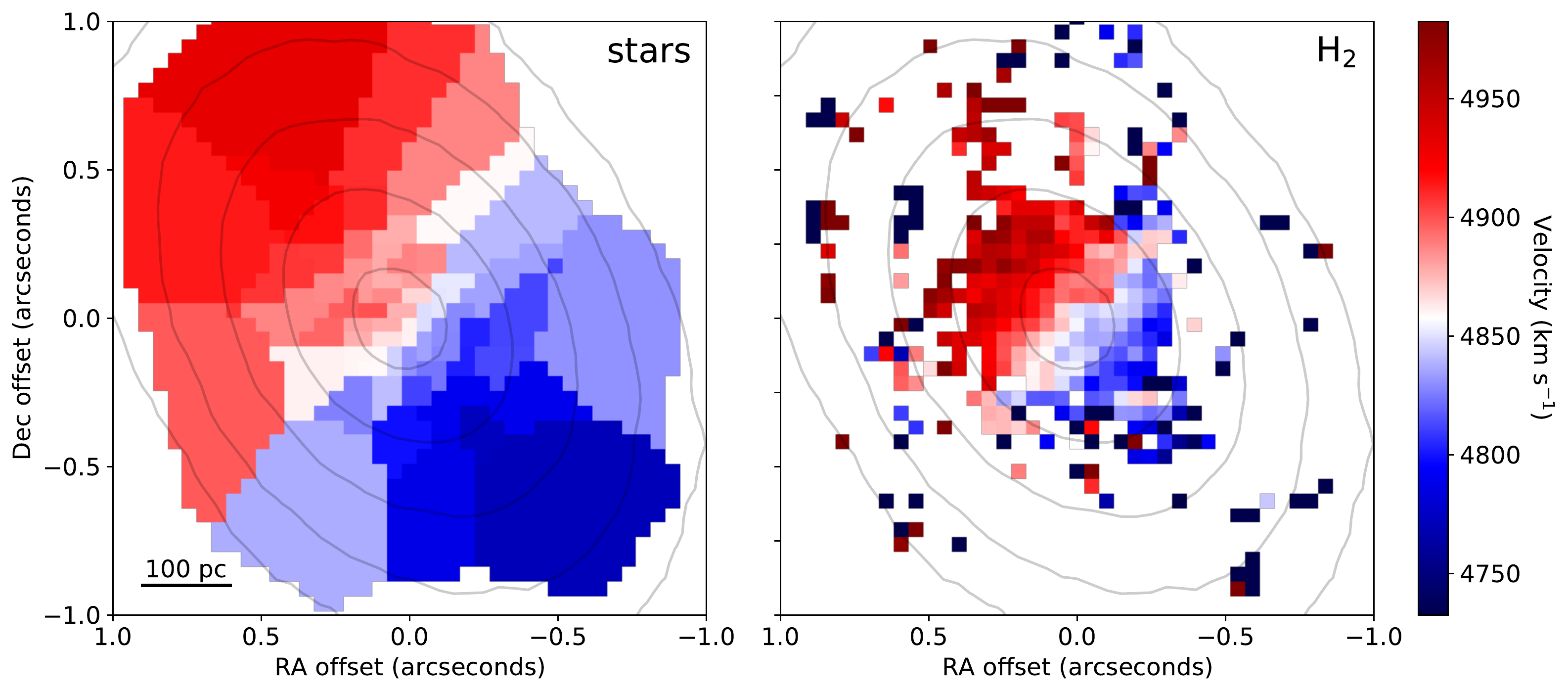}
	\caption{Velocity maps derived from the Gemini NIFS data within the central $2^{\prime \prime} \times 2^{\prime \prime}$ region.  \textit{Left}: Velocity map for the stellar component, using Voronoi binning such that each bin has a signal-to-noise ratio of at least 25.  \textit{Right}: H$_2$ velocity map.  In both panels, mean continuum contours are overplotted at 5\%, 10\%, 20\%, 40\%, and 80\% of the peak value.}
	\label{fig:NIFS_data}
\end{figure*}

\section{Analysis} \label{sec:Analysis}

In this section we describe the analysis procedures used to measure velocities from the Arecibo spectrum (\autoref{sec:HImodeling}) and the Gemini spectra (\autoref{sec:NIFSVelocities}).

\subsection{Neutral hydrogen spectral decomposition and velocity measurements} \label{sec:HImodeling}

Instead of the classic symmetric ``double-horn'' \HI profile \citep{Roberts_1978}, \j0437 shows a more unusual triple-peaked and asymmetric spectral structure.  Because our observations do not spatially resolve the \HI kinematics, the association of individual spectral properties with distinct dynamical components is ambiguous.  While it is clear that a single double-horn component cannot describe the observed spectral profile, there are a variety of more complicated models that could potentially do so adequately.  In \autoref{app:GaussianModeling} we explore three plausible model extensions, from which we conclude that the observed spectral structure is most conservatively and satisfactorily modeled using a sum of Gaussian components.  Using the \dynesty nested sampling routine \citep{Speagle_2020} to explore the posterior distribution, we find that $N=3$ Gaussian components are sufficient to capture the spectral structure and achieve a reduced-$\chi^2$ of $\sim$1 (see \autoref{fig:HI_spectrum}); the velocities for these components are reported in \autoref{tab:Velocities} along with their statistical uncertainties.  Our modeling procedure is described in more detail in \autoref{app:GaussianModeling}.

We use the modeled \HI spectrum to make a measurement of $V_{20}$, defined to be to the midpoint between the two points on the profile that rise to 20\% of the peak amplitude \citep[see, e.g.,][]{Fouque_1990}.  $V_{20}$ provides an estimate of the galaxy recession velocity, and we find $V_{20} = 4909.9 \pm 1.9$\,\kms.  For the associated width of the profile, $W_{20}$, we find $W_{20} = 326.0 \pm 4.2$\,\kms, and for the total mass of \HI we find $M_{\text{\HI}} = (1.35 \pm 0.18) \times 10^8$\,M$_{\odot}$.

\begin{deluxetable*}{lcccc}
\tablecolumns{5}
\tablewidth{0pt}
\tablecaption{Velocity measurements for different components in \j0437\label{tab:Velocities}}
\tablehead{\colhead{\textbf{Source of velocity}} & \colhead{\textbf{Velocity (\kms)}} &  \colhead{\textbf{Uncertainty (\kms)}} & \colhead{\textbf{Spatial scale (pc)}} & \colhead{\textbf{Reference}}}
\startdata
Maser rotation curve & 4818.0 & 10.5 & 0.2 & \citetalias{Gao_2017} \\
Maser rotation curve re-analysis & 4809.3 & 10.0 & 0.2 & this work \\
\midrule
NIFS stellar & 4857--4844 & $\sim$2\hphantom{$\sim$} & 34--340 & this work \\
NIFS H$_2$ & 4858--4875 & $\sim$2\hphantom{$\sim$} & 34--170 & this work \\
NIFS integrated light & 4865--4853 & $\sim$2--4\hphantom{$\sim$} & 45--470 & this work \\
\midrule
SDSS spectrum, emission lines & 4882.2 & 7.7 & \multirow{3}{*}{$1.0 \times 10^3$} & \cite{Pesce_2018} \\
SDSS spectrum, stellar & 4921.4 & 19.1 & & \cite{Pesce_2018} \\
SDSS spectrum, average & 4887.6 & 7.1 & & \cite{Pesce_2018} \\
\midrule
\HI spectrum, first component & 4774.6 & 3.2 & \multirow{4}{*}{$4.9 \times 10^4$} & this work \\
\HI spectrum, second component & 4870.2 & 0.8 &  & this work \\
\HI spectrum, third component & 4989.4 & 1.9 &  & this work \\
\HI spectrum, $V_{20}$ & 4909.9 & 1.9 &  & this work
\enddata
\tablecomments{Velocity measurements considered in this paper and the spatial scales on which they are measured, assuming a distance of 70\,Mpc to \j0437.  For the NIFS velocities, we quote the range of values corresponding to the systemic velocities at the innermost and outermost annuli in which measurements were made; note that for the NIFS stellar measurements, the systemic velocity measured from the outer annulus is smaller than that measured from the inner annulus.  For the NIFS stellar and H$_2$ velocity measurements, the uncertainties are dominated by an overall calibration systematic of $\sim$2\,\kms.}
\end{deluxetable*}

\subsection{Systemic velocities of the stellar and H$_2$ components} \label{sec:NIFSVelocities}

The NIFS $K$-band spectra show both the strong stellar absorption lines of CO at $\sim$2.3\,$\mu$m and molecular hydrogen emission lines, including the strong H$_2$\,1-0\,S(1) line at rest wavelength 2.12\,$\mu$m.

Stellar kinematics were derived by first Voronoi binning the data cube to $S/N \geq 25$ \citep{Cappellari_03}, and then fitting the data with pPXF \citep{Cappellari_04} using high resolution stellar templates from \citet{Wallace_96}.  The resulting radial velocity map can be seen in \autoref{fig:NIFS_data}.  Errors on individual bins are determined through Monte Carlo simulations and range from 5--10\,\kms.  The velocity map was then analyzed using the Kinemetry code \citep{Krajnovic_2006} to determine the barycentric systemic velocity as a function of radius from the observed photocenter; we note that the appearance of the galaxy in the NIFS data cubes is very symmetric.  At the smallest radius (0$\farcs$05), the systemic velocity is 4856.8$\pm$1.6\,\kms.  Kinemetry reveals that the velocity steadily declines with radius -- at 0$\farcs$5 it is 4844.5$\pm$0.6\,\kms.  These measurements are shown as orange dots in \autoref{fig:dist_vs_vel}.  We note that the quoted errors are the formal errors produced by the Kinemetry code and are smaller than the systematic errors in our velocities discussed below.

To check the veracity of the systemic velocity shift with radius, we also binned the spectra in circular annular bins, and we ran pPXF on the resulting spectra.  The mean velocity of the innermost spectrum ($<$0$\farcs$1) is 4865.4$\pm$2.9\,\kms, while the annulus between 0$\farcs$4 and 0$\farcs$6 has a velocity of 4848.8$\pm$3.1\,\kms.  Thus it seems quite clear that there is indeed a blueshift in the systemic velocity of $\sim$15\,\kms between the center of the galaxy and the galaxy at radii of a few hundred parsecs.  These ``integrated light'' measurements are shown as red points in \autoref{fig:dist_vs_vel}.

We also determine the kinematics of H$_2$\,1-0\,S(1) emission line.  Because the emission line strength does not follow the stellar emission distribution, for this measurement we do not bin the data, and instead we measure the velocities of the emission lines in each pixel with a Gaussian fit.  We fit only lines where the total flux in our fitting region is $>$10 times the surrounding noise level.  The result is shown in the right panel of \autoref{fig:NIFS_data}. Clear rotation with the same position angle as the stellar kinematics is visible.  However, Kinemetry reveals that while the systemic velocities of the stellar and H$_2$ components are similar at the innermost radii, at larger radii the H$_2$ systemic velocity is actually redshifted (not blueshifted like the stellar kinematics), as shown by the blue points in \autoref{fig:dist_vs_vel}.

We note that the wavelength solution of the NIFS data was verified through fitting of sky lines in the spectra; a standard deviation of 1.1~km/s from the mean velocity was found from pixel to pixel, and an overall offset of -0.8 km/s was found.  Thus the systematic errors on our velocity measurements are $<$2 km/s, much smaller than the velocity gradient observed in the stellar and gas kinematics.

\subsection{The velocity of the SMBH}\label{sec:SMBH_velocity}

The velocity of the SMBH in \j0437 has previously been measured to be $4818 \pm 10.5$\,\kms by \citetalias{Gao_2017}, who used VLBI measurements of H$_2$O megamasers in the SMBH accretion disk to map out its rotation curve well within the gravitational sphere of influence.  By fitting this rotation curve with a thin-disk Keplerian model, \citetalias{Gao_2017} were able to measure both the mass and velocity of the central SMBH.  In this section, we re-analyze the same VLBI dataset using an updated maser disk model, which relaxes several of the assumptions made by \citetalias{Gao_2017} and thus permits an improved assessment of the associated velocity uncertainty.

The VLBI observations carried out by \citetalias{Gao_2017} resulted in position and velocity measurements for each of the detected maser features, or ``spots.''  \citetalias{Gao_2017} fit their rotation curve using a two-step procedure, in which the entire VLBI map is first rotated and shifted such that the blueshifted and redshifted masers lie on the horizontal axis, and then the maser spot velocities are fit as a function of their measured one-dimensional positions along this horizontal axis.  The \citetalias{Gao_2017} rotation curve model contains three free parameters: the SMBH mass, the one-dimensional SMBH position along the horizontal axis, and the SMBH's line-of-sight velocity.

For the present analysis we employ a modified version of the maser disk model described in \cite{Pesce_2020} to fit the \j0437 VLBI data.  The primary modification is the removal of acceleration measurements from the model likelihood, as the available VLBI dataset does not contain any such acceleration measurements.  Our fitting approach differs from that of \citetalias{Gao_2017} in several respects:
\begin{enumerate}
    \item We take the maser velocities, rather than their positions, to be the ``independent'' quantities; i.e., the model is essentially $r(v)$ rather than $v(r)$.  This strategy leverages the fact that the individual velocity measurements are uncertain at a level comparable to a spectral channel width (${\sim}$1--2\,\kms) and therefore much smaller than the orbital velocities of several hundred \kms, while the position uncertainties are comparatively large ($\sim$0.01--0.1\,mas) relative to the orbital radii of several tenths of a milliarcsecond.
    \item We do not perform any pre-rotation of the VLBI map, and instead we fit for the two-dimensional location of the SMBH on the sky along with the position angle of the disk.
    \item We permit the disk inclination angle to be a free parameter in the fit.
    \item We permit a warp in the position angle of the disk with radius.
    \item We fit for systematic ``error floor'' parameters in the $x$ and $y$ maser position measurements alongside the disk model parameters.  These error floor parameters describe the additional uncertainty that it would be necessary to add into the measurements to ensure that the data are consistent with the model; i.e., these parameters enforce a final reduced-$\chi^2$ value that is consistent with unity.
\end{enumerate}
Detailed descriptions of the model parameters, likelihood, and fitting procedure are provided in \cite{Pesce_2020}.  Following \citetalias{Gao_2017}, we fit only to the redshifted and blueshifted maser features because there are no available acceleration measurements to constrain the systemic maser feature orbital radii.  The final model contains 9 parameters, which are listed in \autoref{tab:MaserModeling} along with their priors and best-fit values.

\begin{deluxetable}{LlCC}[t]
\tablecolumns{4}
\tablewidth{0pt}
\tablecaption{Results from re-analysis of maser VLBI data \label{tab:MaserModeling}}
\tablehead{\colhead{Parameter} & \colhead{Units} & \colhead{Prior} & \colhead{Best-fit value}}
\startdata
v_0 & \kms & \mathcal{U}(4500,5500) & 4809.3 \pm 10.0 \\
M & $10^6$\,$\text{M}_{\odot}$ & \mathcal{U}(0,30) & 2.86 \pm 0.2 \\
x_0 & mas & \mathcal{U}(-0.5,0.5) & 0.096 \pm 0.005 \\
y_0 & mas & \mathcal{U}(-0.5,0.5) & 0.109 \pm 0.014 \\
i_0 & deg. & \mathcal{U}(70,110) & \text{unconstrained} \\
\Omega_0 & deg. & \mathcal{U}(0,180) & 16.9 \pm 2.5 \\
\Omega_1 & deg.\,mas$^{-1}$ & \mathcal{U}(-100,100) & 14 \pm 8 \\
\sigma_x & $\mu$as & \mathcal{U}(0,1000) & 6 \pm 1.5 \\
\sigma_y & $\mu$as & \mathcal{U}(0,1000) & <5 \\
\enddata
\tablecomments{Results from fitting a thin Keplerian disk model to the \j0437 VLBI maser dataset from \citetalias{Gao_2017}, as described in \autoref{sec:SMBH_velocity} and shown in \autoref{fig:maser_rotation_curve}.  The fitted model parameters are the SMBH velocity $v_0$, the SMBH mass $M$, the SMBH position $(x_0,y_0)$, the disk inclination $i_0$, the disk position angle $\Omega_0$, a first-order warp in the disk position angle with radius $\Omega_1$, and two error floor parameters $\sigma_x$ and $\sigma_y$ for the maser $x$- and $y$-position measurements, respectively.  The notation $\mathcal{U}(a,b)$ denotes a uniform distribution on the range $(a,b)$.  For most parameters we report the posterior mean and standard deviation, though we note that the $\sigma_y$ parameter has a best-fit value that is consistent with zero and so we report the 95\% upper limit instead.  For the $i_0$ parameter, the posterior distribution matches the prior distribution, and so we do not report constraints on this parameter.  A more detailed description of the various model parameters can be found in \cite{Pesce_2020}.}
\end{deluxetable}

\begin{figure*}[t]
	\centering
		\includegraphics[width=1.00\textwidth]{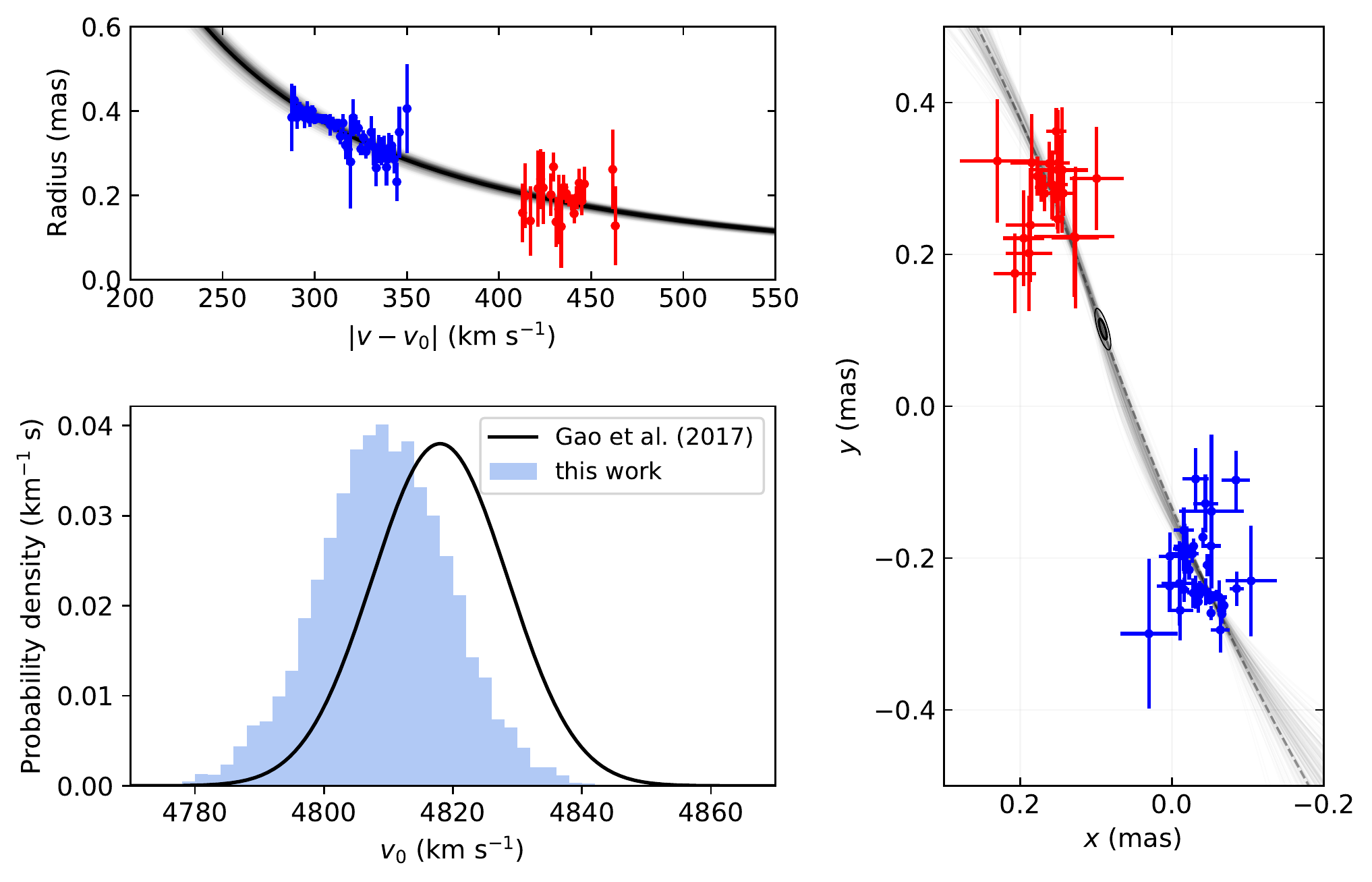}
	\caption{Results from fitting a thin Keplerian disk model to the \j0437 maser measurements from \citetalias{Gao_2017}; the best-fit model parameters are listed in \autoref{tab:MaserModeling}.  The top left panel shows the on-sky projected radial separation from the SMBH versus orbital velocity for each of the maser spots, with the best-fit rotation curve plotted in black and 200 draws from the posterior distribution plotted in gray.  The points corresponding to individual maser features have been colored by velocity group, with blue points denoting blueshifted maser features and red points denoting redshifted maser features.  The bottom left panel shows the posterior distribution for the SMBH velocity that we obtain from our fitting procedure (blue histogram) along with a Gaussian distribution with the mean and standard deviation reported in \citetalias{Gao_2017} (black line). The right panel shows the VLBI map of the maser system, with the best-fit warped disk midplane plotted as a dashed line and 200 draws from the posterior distribution plotted as solid gray lines; the 1$\sigma$ and 2$\sigma$ contours for the SMBH location are shown as thick and thin black ellipses, respectively.}
	\label{fig:maser_rotation_curve}
\end{figure*}

Our best-fit rotation curve and disk model are shown in \autoref{fig:maser_rotation_curve}, from which we determine the SMBH velocity to be $4809.3 \pm 10.0$\,\kms.  We find that the uncertainty in the derived velocity matches well with the result from \citetalias{Gao_2017}, though our best-fit velocity itself is approximately 9\,\kms smaller.  Because our disk model relies on fewer assumptions than that employed in \citetalias{Gao_2017}, we hereafter adopt $4809.3 \pm 10.0$\,\kms as the velocity measurement of the SMBH in \j0437.

\section{Discussion} \label{sec:Discussion}

\begin{figure}[t]
	\centering
		\includegraphics[width=1.00\columnwidth]{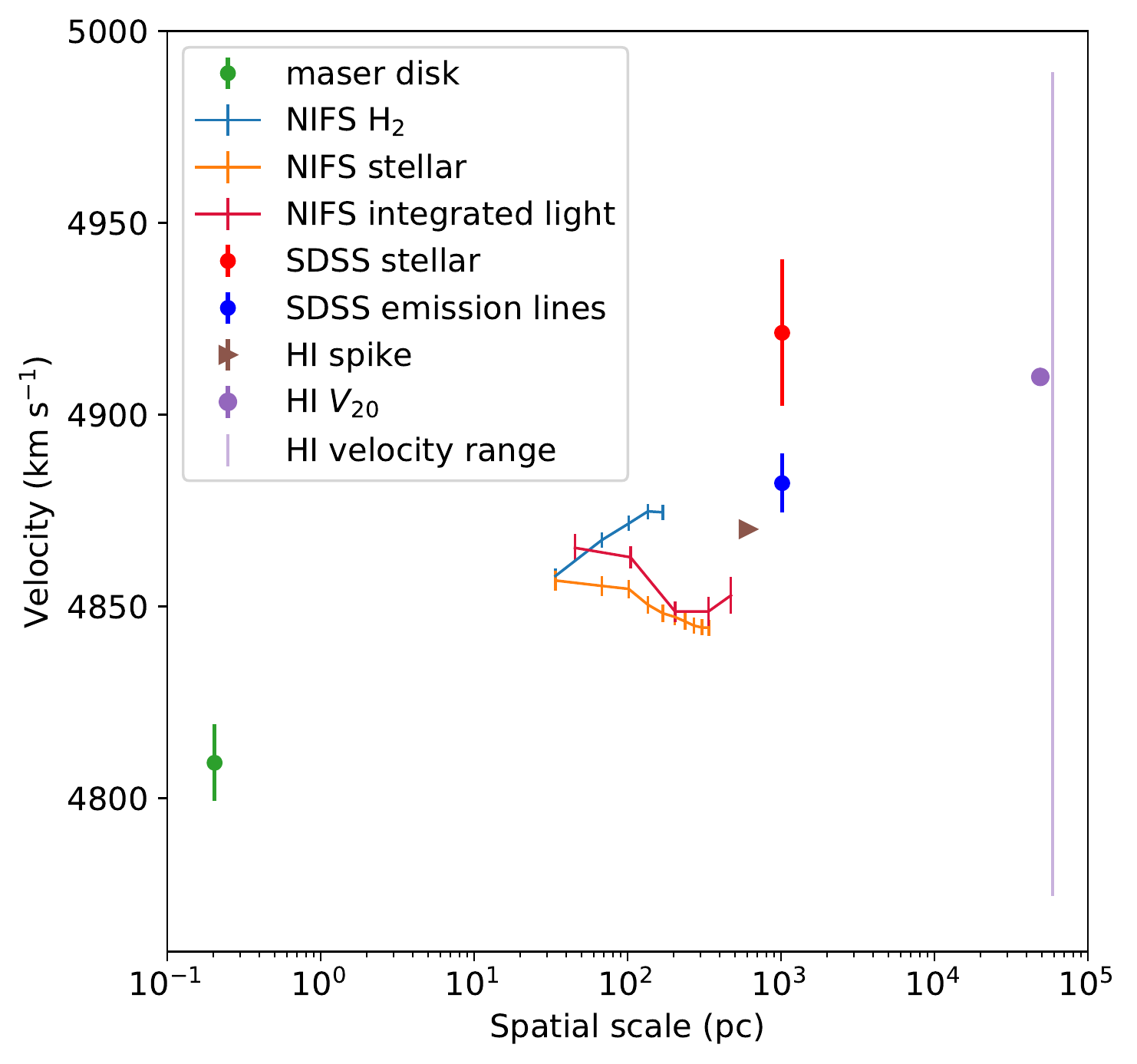}
	\caption{The spatial scales on which the various velocity measurements considered in this paper are made.  For the \HI spike, we take the spatial scale to be $\geq$600\,pc as implied by the brightness temperature limit given in \autoref{eqn:SpinTempLimit}.  The full peak-to-peak velocity range spanned by the \HI emission (corresponding to the horizontal green line at the bottom of \autoref{fig:HI_spectrum}) is shown as a vertical line that is horizontally offset from the $V_{20}$ velocity for visual clarity. For the NIFS measurements, we plot the systemic velocities as a function of annulus diameter and we include an overall 2\,\kms calibration systematic uncertainty on the error bars.  All other velocities are plotted with statistical error bars.}
	\label{fig:dist_vs_vel}
\end{figure}

The recession velocity measurements considered in this paper are listed in \autoref{tab:Velocities}, and they are plotted against spatial scale in \autoref{fig:dist_vs_vel}.  We find that all velocity measurements fall within a $\sim$100\,\kms range spanning ${\sim}$4820--4920\,\kms, and there is a general trend for measurements made at larger spatial scales to recover larger recession velocities.  In this section we discuss the various measurements and consider some possible interpretations.

\subsection{Velocity measurements in \j0437}\label{sec:VelMeasurements}

The largest spatial scales are probed by the \HI emission, which traces gas throughout the galaxy and out to the edge of the Arecibo beam (roughly $\sim$50\,kpc across).  \j0437's \HI profile is atypical in that it shows three prominent spectral peaks rather than the usual two that are expected for a simply-rotating system.  Similar profiles have been classified as ``anomalous'' by previous authors \citep[e.g., UGC 2889 in ][]{Courtois_2009}, and they are often attributed to spatial blending of galaxy pairs in single-dish spectra, such as in the case of NGC 876 and NGC 877 \citep{Bottinelli_1982,Lee-Waddell_2014}.  However, for \j0437 we see neither evidence for a companion galaxy within the Arecibo beam (see \autoref{fig:SDSS_image}) nor obvious signs of morphological disturbance in Hubble Space Telescope (HST) images \citep{Pjanka_2017}.  Nevertheless, the measured \HI central velocity of $V_{20} = 4910$\,\kms is in agreement with the SDSS stellar velocity measured by \citetalias{Pesce_2018}, supporting the notion that both measurements trace the recession velocity of \j0437.  Furthermore, the \HI central velocity is in ${\sim}10{\sigma}$ disagreement with the SMBH velocity as measured from the maser rotation curve (\autoref{sec:SMBH_velocity}), indicating that the black hole is blueshifted by roughly 100\,\kms with respect to the galaxy's recession velocity.

The NIFS measurements probe spatial scales of $\sim$30--300\,pc.  The sense of rotation for both the stellar and H$_2$ components agrees with that of the maser disk \citepalias{Gao_2017}, though the maser disk has a position angle of ${\sim}20^{\circ}$ while the outermost stellar and H$_2$ components have position angles of ${\sim}40^{\circ}$.\footnote{We note that this ${\sim}20^{\circ}$ difference in position angle is consistent with the offsets between maser disks and circumnuclear structures seen in other galaxies \citep{Greene_2013} and comparable to the ${\sim}16^{\circ}$ position angle difference between the \j0437 maser disk and nuclear structure reported in \cite{Pjanka_2017}.}  We find that the systemic velocities of the stellar and H$_2$ rotation curves agree with one another on the smallest scales ($\sim$30\,pc), though they both show a $\sim$4.7$\sigma$ redshift with the respect to the maser velocity.  At larger radii the stellar and H$_2$ systemic velocity measurements diverge, with the stellar systemic velocity showing a $\sim$30\,\kms blueshift with respect to the H$_2$ on scales of $\sim$200\,pc.  Such large variations in the measured stellar systemic velocity as a function of radius are rare; the typical dispersion of \atlas galaxies between the central and $r=500$\,pc velocities is only $\sim$3--4\,\kms (\autoref{app:ATLAS3DSysVel}; \citealt{Krajnovic_2011}), and most of the galaxies with substantially larger systemic velocity gradients show evidence of interaction.  We note, however, that such a relative velocity offset could also be plausibly explained by a combination of geometric and obscuration effects (e.g., if the stellar and $H_2$ emission arose from two separate misaligned and mutually obscuring disks of material) while leaving the system dynamically relaxed, and that the structure maps produced by \cite{Pjanka_2017} do show evidence of dust on $\sim$0.3$^{\prime \prime}$ and larger scales.

The outermost H$_2$ emission ($\sim$200\,pc) has a systemic velocity of 4875\,\kms that matches well with the emission line velocity measured by \citetalias{Pesce_2018} from the SDSS spectrum, indicating that these two measurements may be tracing similar material.  These measurements are both also in agreement with the velocity of the ``anomalous'' central \HI spike, which has a velocity of $\sim$4870\,\kms and an amplitude of $S_{\nu} \approx 0.7$\,mJy (see \autoref{tab:HISpectralModeling}).  If this \HI spike represents a distinct dynamical subsystem (rather than, e.g., one ``horn'' of a double-horn profile), then we can set a lower limit on the area of the emission region by requiring that the \HI brightness temperature not exceed its spin temperature of $T_s \approx 150$\,K \citep{Condon_2016},

\begin{equation}
\Omega \geq \frac{S_{\nu} c^2}{2 k \nu^2 T_s} . \label{eqn:SpinTempLimit}
\end{equation}

\noindent Here, $k$ is the Boltzmann constant, $\nu = 1.4$\,GHz is the emitting frequency, and $\Omega$ is the solid angle subtended by the emitting region.  \autoref{eqn:SpinTempLimit} implies that the angular size of the region contributing the \HI spike is ${\sim}1.8^{\prime \prime} \approx 600$\,pc.  This spatial scale is similar to that probed by the NIFS observations, and together with the coincident velocities suggests that all three sources of emission -- i.e., the outermost H$_2$, the SDSS emission lines, and the \HI spike -- may be originating from material with shared dynamics.  The velocity of this material is significantly different from that of both the SDSS stellar and the central \HI velocity (i.e., $V_{20}$), perhaps indicating that there is a kinematically distinct subsystem located in the centermost few hundred parsecs of \j0437.  However, we note that the observed FWHM of the \HI spike of only $\sim$55\,\kms (see \autoref{tab:HISpectralModeling}) is in tension with this interpretation, because at several-hundred parsec radii the material in this galaxy should display a FWHM of $\sim$200\,\kms (\autoref{fig:NIFS_data}; see also \citealt{Noordermeer_2007}).  It thus may not be viable to interpret this \HI spike as a distinct kinematic component.

\subsection{Uncertainty in the black hole velocity measurement}

Our measurement of the SMBH velocity (see \autoref{sec:SMBH_velocity}) relies on accurate VLBI position measurements for each of the maser features, and if there are unaccounted-for systematic uncertainties in these position measurements then we would expect the velocity measurement to be correspondingly impacted.  \citetalias{Gao_2017} considered the impact of phase referencing uncertainties on the \j0437 maser position measurements.  The absolute sky location of the \j0437 peak maser emission (i.e., the emission at a velocity of 4505.8\,\kms used as a reference feature) is known from phase-referenced VLBI measurements to a precision of better than 2\,mas.  \citetalias{Gao_2017} estimate that the expected additional positional uncertainties associated with this imperfectly-known reference position, when propagated to the rest of the maser features, should be ${\lesssim}$5\,$\mu$as.  This expectation is consistent with the magnitudes of the error floor parameters that we recover from our model fitting (see \autoref{tab:MaserModeling}).  Additionally, we note that there are no obvious systematic trends in the residual dispersion about the best fit such as would be expected if poor phase calibration were present at this level.

\subsection{An offset black hole}

In our own Galactic Center, we have high-precision evidence that the SMBH is coincident with the dynamical center of the Galaxy \citep{Reid_2020}.  While we believe that a similar situation should generally hold for other galaxies as well, a number of effects can at least temporarily knock the SMBH out of this equilibrium position.  At very low galaxy mass, it is possible that SMBHs never settle at their galaxy center, given the very shallow galactic potential \citep[e.g.,][]{Bellovary_2019,Reines_2020}. However, at higher galaxy masses, it is most likely that mergers are responsible for SMBH motions.

Relative motions and spatial offsets between SMBHs and their host galaxies occur throughout the merger process.  As galaxies merge, the SMBHs from each galaxy will be offset both spatially and in velocity from the center of the merger.  This stage may be observable as velocity offset active galaxies \citep[e.g.,][]{Comerford_2009,Comerford_2014} or as spatially resolved pairs of active galactic nuclei (AGN; e.g., \citealt{Komossa_2003,Gerke_2007}). Further along in the merger process, when the two SMBHs become gravitationally bound, one may hope to observe the signatures of orbital motion for the bound pair \citep[e.g.,][see also \autoref{app:Binary}]{Eracleous_2012,Shen_2013,Ju_2013}. Finally, if an SMBH merger occurs, then any anisotropy in the radiated linear momentum will lead to a gravitational wave recoil \citep{Fitchett_1983}.  These have been many observational recoil candidates proposed, but all have their complications (see reviews in \citealt{Komossa_2012} and \citealt{Blecha_2016}).

The SMBH in the galaxy \j0437 is, to our knowledge, the most concrete case of an SMBH in motion with respect to its galaxy.  Because our initial search focused on megamaser disk galaxies \citepalias{Pesce_2018}, the sources were all within 200\,Mpc where detailed followup observations are possible; luminous AGN that have been identified as recoil or binary SMBH candidates in the past are often much more distant.  Even in the case of \j0437, ambiguity remains about whether we are seeing an SMBH making its way to the galaxy center for the first time, SMBH binary orbital motion, or a recoil product.  However, the fact that the galaxy on large scales is apparently out of equilibrium provides indirect evidence that we are observing the aftermath of a merger.

\section{Summary and conclusion} \label{sec:Conclusion}

Following the identification in \citetalias{Pesce_2018} of the galaxy \j0437 as a candidate for hosting a binary or recoiling SMBH, we have obtained Arecibo and Gemini NIFS observations of the galaxy.  Our new observations support the claim of a velocity offset between the SMBH and its host galaxy.  Furthermore, the systemic velocity in \j0437 exhibits an apparent spatial scale dependence; the overall picture looks something like the following:

\begin{enumerate}
    \item On the smallest spatial scales ($<$1\,pc), where the motion of gas is dominated by the gravitational potential of the SMBH, H$_2$O masers orbit with a central velocity of $\sim$4810\,\kms.  We associate this velocity with the SMBH itself.
    \item At the photocenter of the galaxy, within the central $\sim$30\,pc and coincident with the location of the SMBH, both the stars and H$_2$ gas emission lines have a systemic velocity of $\sim$4860\,\kms.  However, on somewhat larger scales ($\sim$30--200\,pc), both gas and stars exhibit unusual systemic velocity gradients of $\sim$15\,\kms in opposite directions.  In all cases, these velocities are significantly offset from the SMBH velocity as traced by the masers.
    \item On the largest spatial scales ($\sim$1--10\,kpc), the velocity of the \HI emission is in agreement with the SDSS stellar velocity from \citetalias{Pesce_2018}.  We find a central \HI velocity of $V_{20} \approx 4910$\,\kms that we associate with the recession velocity of the galaxy as a whole, though we note that the ``anomalous'' structure of the \HI spectral profile complicates this interpretation.
\end{enumerate}

Multiple lines of evidence -- including the different inferred systemic velocities on different spatial scales, the ``anomalous'' \HI spectral structure, and the gradient in stellar systemic velocity with radius -- point to the conclusion that the galaxy \j0437 has been dynamically perturbed sometime in the recent past, likely through an interaction with another galaxy.  Of particular interest is the apparent difference between the systemic velocity of the SMBH and that of any other dynamical tracer, indicating that the SMBH in this galaxy is in motion with respect to the surrounding material.

\citetalias{Pesce_2018} explored plausible causes of such relative motion, ultimately settling on three possibilities: (1) the SMBH originates from an external galaxy that is in the process of merging with \j0437; (2) the SMBH is part of a binary system, and the velocity offset we observe is the result of its orbital motion; or (3) the SMBH is recoiling from a recent merger event.  Though any of these possibilities would be exciting, with the current data we are unfortunately unable to distinguish between them.  Additional observations are required to ascertain the nature of the peculiar SMBH in \j0437.

\acknowledgments

We are grateful to Robert Minchin, Joan Schmelz, and Arun Venkataraman at Arecibo Observatory for their help with data acquisition and reduction.

This paper makes use of observations taken using the Gemini Observatory under program GN-2018B-FT-110 and the Arecibo Observatory under programs A3241 and A3300.  The Arecibo Observatory is operated by SRI International under a cooperative agreement with the National Science Foundation (AST-1100968), and in alliance with Ana G. M\'endez-Universidad Metropolitana, and the Universities Space Research Association.  Support for this work was provided by the NSF through grants AST-1952099, AST-1935980, AST-1828513, and AST-1440254, and by the Gordon and Betty Moore Foundation through grant GBMF-5278.  This work was supported in part by the Black Hole Initiative at Harvard University, which is funded by grants from the John Templeton Foundation and the Gordon and Betty Moore Foundation to Harvard University.  ACS acknowledges support from NSF AST-1350389.

This research made use of Montage, which is funded by the National Science Foundation under Grant Number ACI-1440620, and was previously funded by the National Aeronautics and Space Administration's Earth Science Technology Office, Computation Technologies Project, under Cooperative Agreement Number NCC5-626 between NASA and the California Institute of Technology.

Funding for the SDSS and SDSS-II has been provided by the Alfred P. Sloan Foundation, the Participating Institutions, the National Science Foundation, the U.S. Department of Energy, the National Aeronautics and Space Administration, the Japanese Monbukagakusho, the Max Planck Society, and the Higher Education Funding Council for England. The SDSS Web Site is \url{http://www.sdss.org/}.

The SDSS is managed by the Astrophysical Research Consortium for the Participating Institutions. The Participating Institutions are the American Museum of Natural History, Astrophysical Institute Potsdam, University of Basel, University of Cambridge, Case Western Reserve University, University of Chicago, Drexel University, Fermilab, the Institute for Advanced Study, the Japan Participation Group, Johns Hopkins University, the Joint Institute for Nuclear Astrophysics, the Kavli Institute for Particle Astrophysics and Cosmology, the Korean Scientist Group, the Chinese Academy of Sciences (LAMOST), Los Alamos National Laboratory, the Max-Planck-Institute f\"ur Astronomy (MPIA), the Max-Planck-Institute f\"ur Astrophysics (MPA), New Mexico State University, Ohio State University, University of Pittsburgh, University of Portsmouth, Princeton University, the United States Naval Observatory, and the University of Washington.

\facilities{Arecibo Observatory, Gemini North} 
\software{AOIDL, \texttt{dynesty} \citep{Speagle_2020}, Montage\footnote{\url{http://montage.ipac.caltech.edu}}, Gemini IRAF, Kinemetry \citep{Krajnovic_2006}, pPXF \citep{Cappellari_04}}

\clearpage
\appendix

\section{\HI spectral modeling} \label{app:GaussianModeling}

Here we describe three different models we use to fit the \HI spectrum from \autoref{sec:HImodeling}.  For each model, we use a Gaussian likelihood given by

\begin{equation}
\ln\left( \mathcal{L} \right) = - \frac{1}{2} \sum_j \left[ \left( \frac{S_{\nu}(v_j) - \hat{S}_{\nu}(v_j)}{\sigma} \right)^2 + \ln\left( 2 \pi \sigma^2 \right) \right] ,
\end{equation}

\noindent where $S_{\nu}(v_j)$ is the model flux density for a spectral channel with velocity $v_j$, $\hat{S}_{\nu}(v_j)$ is the observed flux density in that channel, $\sigma$ is the flux density uncertainty in a single channel, and the sum is taken over all channels.  This likelihood assumes that every spectral channel contains independent Gaussian-distributed noise with a standard deviation $\sigma$ that we treat as a model parameter in each of our fits.  We use the \texttt{dynesty} nested sampling code \citep{Speagle_2020} for posterior exploration.  The best-fit values and uncertainties for all model parameters are listed in \autoref{tab:HISpectralModeling}.

\subsection{Modeling the profile using a sum of Gaussian components} \label{sec:ThreeGaussModeling}

The model we use in our primary analysis (\autoref{sec:HImodeling}) describes the \HI spectral structure using a sum of Gaussian components,

\begin{equation}
S_{\nu}(v) = \sum_{i=1}^N A_i \exp\left[ - \frac{1}{2} \left( \frac{v - v_i}{\sigma_i} \right)^2 \right] , \label{eqn:GaussianSum}
\end{equation}

\noindent where the model parameters are the amplitude $A_i$, central velocity $v_i$, and width $\sigma_i$ for each component.  The total number of model parameters is $3N + 1$, where $N$ is the number of Gaussian components; in this paper, we use $N=3$.  We impose uniform priors on all model parameters, in the range $[0,1]$~mJy for Gaussian component amplitudes, $[0,500]$~\kms for all Gaussian component standard deviations, $[4500,5300]$~\kms for all Gaussian component central velocities, and $[0,1]$~mJy for $\sigma$.  The posterior distribution is trivially multimodal upon pairwise swaps of Gaussian components, but the modes are widely separated in parameter space and so we isolate a single mode when reporting parameter statistics.  A plot of the resulting fit to the spectrum is shown in the left panel of \autoref{fig:HI_spectrum2}.

\begin{deluxetable}{llcC}[t]
\tablecolumns{4}
\tablewidth{0pt}
\tablecaption{Results from \HI spectral modeling \label{tab:HISpectralModeling}}
\tablehead{\colhead{Model description} & \colhead{Parameter description} & \colhead{Units} & \colhead{Best-fit value}}
\startdata
\multirow{10}{5cm}{three Gaussian components} & central velocity of first component & \kms  & 4774.6 \pm 3.2 \\
 & central velocity of second component & \kms  & 4870.2 \pm 0.8 \\
 & central velocity of third component & \kms  & 4989.4 \pm 1.9 \\
 & FWHM of first component & \kms   & 76.4 \pm 6.4 \\
 & FWHM of second component & \kms   & 54.6 \pm 2.3 \\
 & FWHM of third component & \kms   & 129.0 \pm 4.1 \\
 & amplitude of first component     & mJy   & 0.20 \pm 0.01 \\
 & amplitude of second component     & mJy   & 0.66 \pm 0.02 \\
 & amplitude of third component     & mJy  & 0.45 \pm 0.01 \\
 & thermal noise level (RMS) & mJy  & 0.068 \pm 0.002 \\
\midrule
\multirow{13}{5cm}{two double-horn components} & central velocity of first component & \kms & 4810.7 \pm 2.0 \\
 & central velocity of second component & \kms & 4955.9 \pm 2.8 \\
 & velocity width of first component & \kms & 139.6 \pm 4.2 \\
 & velocity width of second component & \kms & 234.0 \pm 9.4 \\
 & flux of first component & mJy\,\kms & 51.7 \pm 3.5 \\
 & flux of second component & mJy\,\kms & 97.6 \pm 2.6 \\
 & asymmetry of first component & unitless & 0.46 \pm 0.09 \\
 & asymmetry of second component & unitless & 0.38 \pm 0.07 \\
 & solid-body fraction of first component & unitless & 0.03 \pm 0.03 \\
 & solid-body fraction of second component & unitless & 0.89 \pm 0.07 \\
 & velocity dispersion of first component & \kms & 12.0 \pm 1.2 \\
 & velocity dispersion of second component & \kms & 11.0 \pm 3.1 \\
 & thermal noise level (RMS) & mJy & 0.064 \pm 0.002 \\
\midrule
\multirow{10}{5cm}{one double-horn component and one Gaussian component} & central velocity of double-horn component & \kms  & 4904.1 \pm 3.7 \\
 & velocity width of double-horn component & \kms & 312.5 \pm 10.2 \\
 & flux of double-horn component & mJy\,\kms & 124.3 \pm 3.2 \\
 & asymmetry of double-horn component & unitless & 0.61 \pm 0.05 \\
 & solid-body fraction of double-horn component & unitless & 0.63 \pm 0.14 \\
 & velocity dispersion of double-horn component & \kms & 18.5 \pm 5.9 \\
 & central velocity of Gaussian component & \kms & 4870.9 \pm 0.7 \\
 & FWHM of Gaussian component & \kms & 38.1 \pm 2.1 \\
 & amplitude of Gaussian component & mJy  & 0.47 \pm 0.02 \\
 & thermal noise level (RMS) & mJy  & 0.066 \pm 0.002 \\
\enddata
\tablecomments{Results from fitting the three different models described in \autoref{app:GaussianModeling} to the Arecibo \HI spectrum; these fits are shown in \autoref{fig:HI_spectrum2}.  For each parameter, we quote the posterior mean and standard deviation.  The thermal noise has been determined per 61\,kHz ($\approx$1.33\,\kms) spectral channel.}
\end{deluxetable}

\subsection{Modeling the profile using a sum of double-horn components} \label{sec:DHmodeling}

Given that the galaxy \j0437 shows signs of dynamical disturbance (potentially indicating a recent merger) and that it exhibits an ``anomalous'' \HI profile (\autoref{fig:HI_spectrum}), it is natural to ask whether a combination of double-horn profiles could give rise to the observed spectral structure.  We have thus performed an alternative analysis using a sum of two double-horn components, each described using the parameterization developed by \cite{Stewart_2014} for each component.

\begin{figure}[t]
	\centering
		\includegraphics[width=1.00\columnwidth]{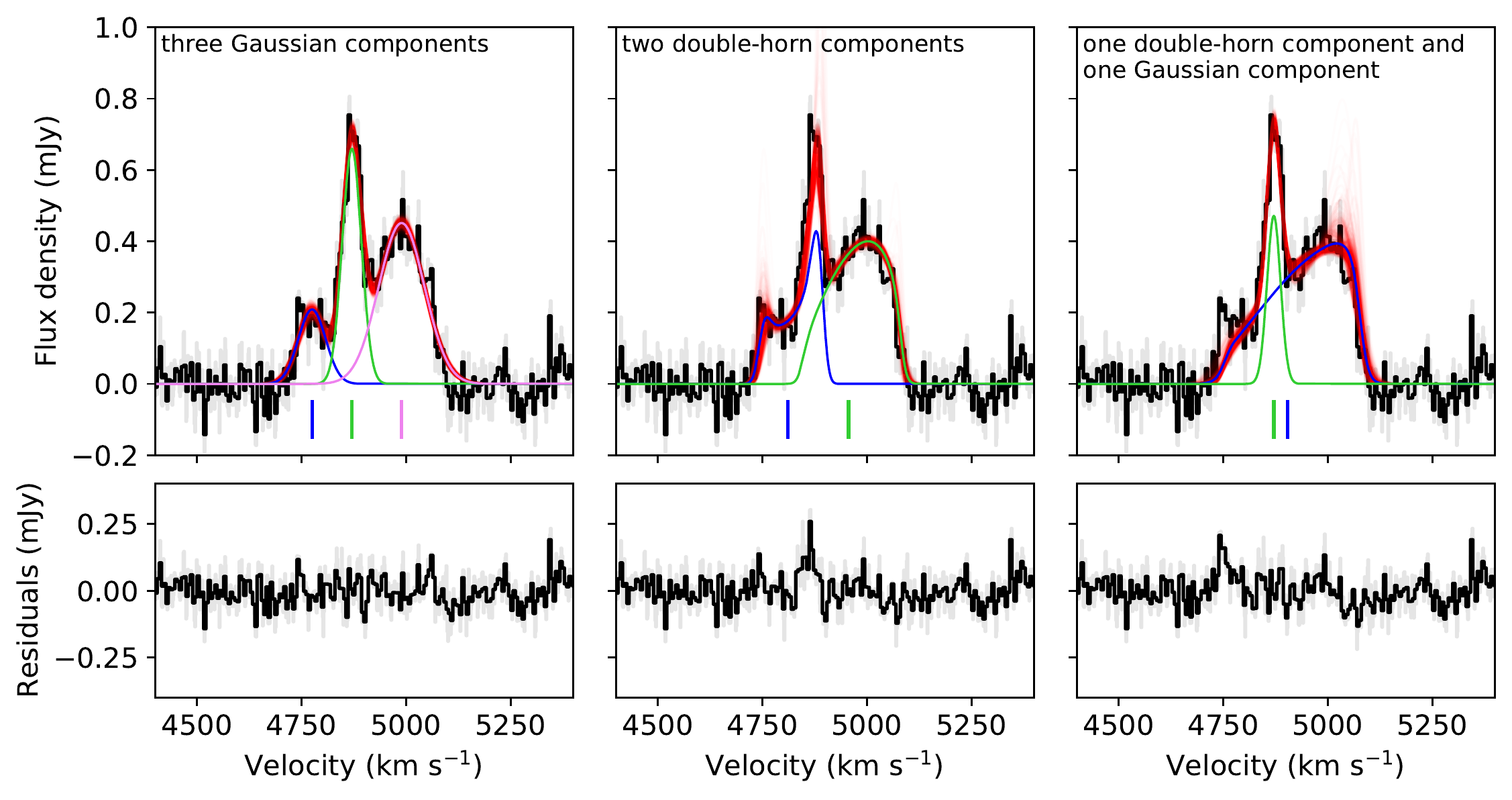}
	\caption{Similar to \autoref{fig:HI_spectrum}, but showing the results of fits using the three different classes of model described in \autoref{app:GaussianModeling}; the best-fit parameter values for each model are listed in \autoref{tab:HISpectralModeling}.  In the left panel we show a more detailed breakdown of the fit from \autoref{fig:HI_spectrum} using three Gaussian components, with the individual best-fit Gaussian model components plotted in blue, green, and violet.  Their corresponding best-fit velocities are marked by the vertical lines underneath each component.  In the center panel we show a similar breakdown for the fit using two double-horn components, and in the right panel the fit using one double-horn component and one Gaussian component.  In all panels, the spectrum is plotted at its native spectral resolution in gray, the spectrum after smoothing by a 4-channel boxcar is shown in black, and 1000 random posterior samples are overplotted in red.  The bottom row of plots shows the residuals (i.e., the difference between the data and best-fitting model) for each fit.}
	\label{fig:HI_spectrum2}
\end{figure}

The \cite{Stewart_2014} model describes a double-horn profile using six parameters: the total flux, the central velocity, the velocity width, an asymmetry parameter, a parameter describing what fraction of the emission comes from solid-body rotation, and a velocity dispersion.  Because we model the spectrum as a sum of $N$ such double-horn components, and because we additionally model the channel uncertainty $\sigma$, the total number of model parameters is $6N+1$; in this paper, we use $N=2$.  We impose uniform priors on all model parameters, in the range $[0,1]$\,Jy\,\kms for the total flux, $[4500,5300]$\,\kms for the central velocity, $[0,600]$\,\kms for the velocity width, $[-1,1]$ for the asymmetry parameter, $[0,1]$ for the solid-body fraction, $[0,100]$\,\kms for the velocity dispersion, and $[0,1]$\,mJy for $\sigma$.

The results of fitting this alternative model to the \HI data are shown in the central panel of \autoref{fig:HI_spectrum2}.  We find that the best-fit model prefers only one of the two components to exhibit a standard double-horn profile, while the other component is dominated by the solid-body contribution and so has only a single, wide spectral peak.  This model struggles to fit the central \HI spike, as evidenced by the large residual flux excess near $\sim$4850\,\kms, so we disfavor it compared to the model composed of three Gaussian components.

\subsection{Modeling the profile using a sum of double-horn and Gaussian components} \label{sec:DHGmodeling}

Motivated by the appearance of the \HI spectrum, we also attempt to model it using a sum of one double-horn component (parameterized as in \citealt{Stewart_2014} and \autoref{sec:DHmodeling}) and one Gaussian component.  The resulting parameter values are listed in \autoref{tab:HISpectralModeling} and the best-fit spectrum is plotted in \autoref{fig:HI_spectrum2}.  We again find that even the best-fit model struggles to fit the observed spectral profile, with a substantial flux excess seen in the residuals around $\sim$4750\,\kms.  We thus disfavor this model compared to the model composed of three Gaussian components.

\section{\atlas systemic velocity curves} \label{app:ATLAS3DSysVel}

The \atlas project has collected integral field spectroscopic measurements for a sample of 260 early-type galaxies in the local Universe \citep{Cappellari_2011}.  This sample provides a reference against which we can gauge the behavior of the NIFS stellar systemic velocity measurements for \j0437, which show a systematic trend with radius (see \autoref{sec:NIFSVelocities}).

\begin{figure*}[t]
	\centering
		\includegraphics[width=1.00\textwidth]{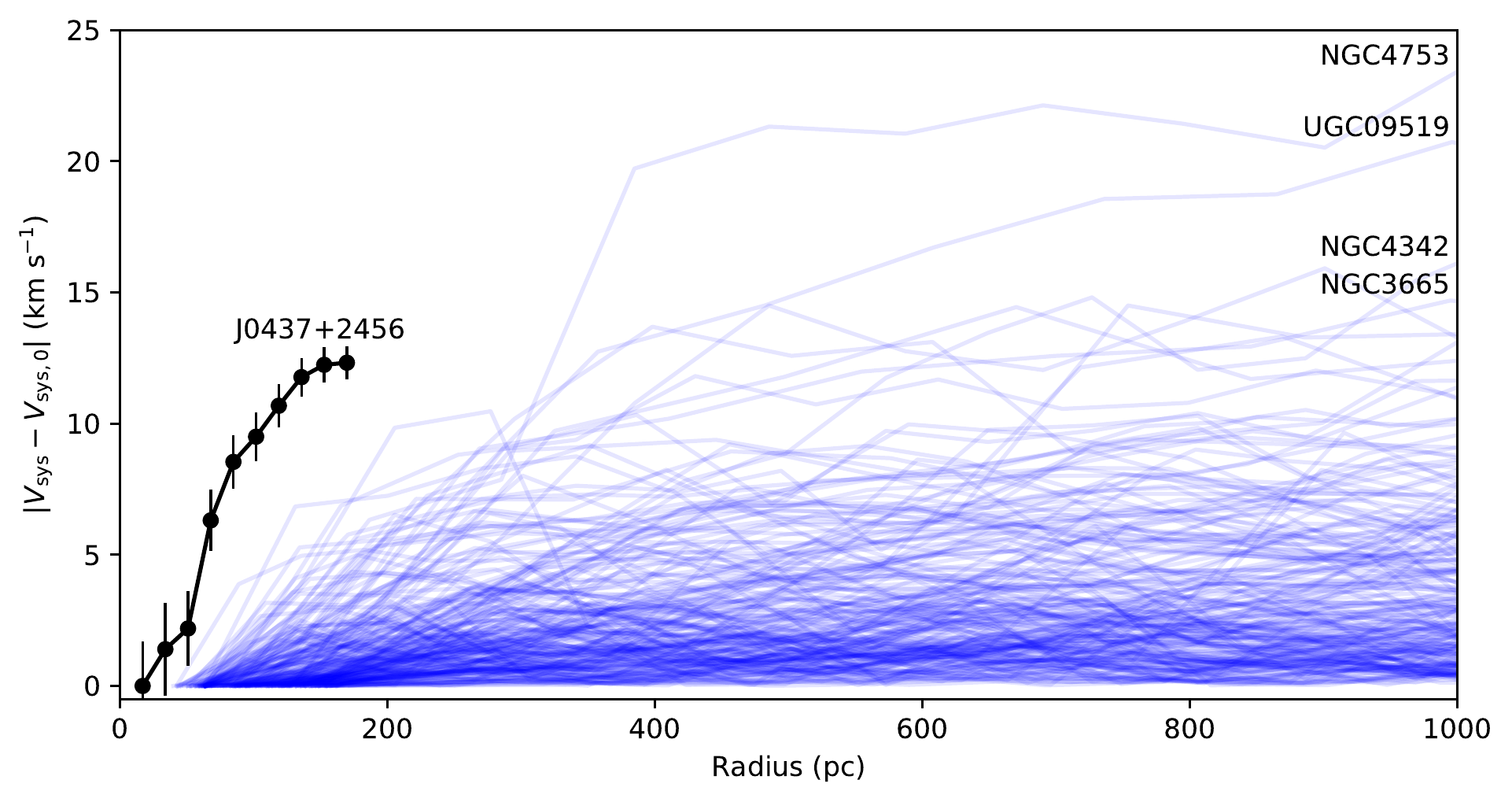}
	\caption{The radial profile of the \j0437 systemic velocity as measured from the NIFS stellar emission (plotted in black; see \autoref{sec:NIFSVelocities}) compared against similar profiles for ``fast rotator'' galaxies from \atlas (plotted in blue).  For each galaxy, we have subtracted off the systemic velocity measured at the smallest radii ($V_{\text{sys},0}$) and then taken an absolute value of the difference to aid comparison.}
	\label{fig:ATLAS3D}
\end{figure*}

\autoref{fig:ATLAS3D} shows the radial profile of the \j0437 stellar systemic velocity measurements plotted alongside the same quantity measured for the ``fast rotator'' galaxies from \atlas.  The \atlas sample is made up of early-type galaxies, while \j0437 is a spiral, so for comparison we select only fast rotators from the \atlas sample because they are galaxies with high angular momentum \citep{2011MNRAS.414..888E}, stellar disks, and ordered (i.e., disk-like) stellar kinematics \citep{Krajnovic_2011, 2013MNRAS.432.1768K}.  We note that unlike the $\sim$0.3-arcsecond seeing of our NIFS observations (see \autoref{sec:Gemini}), many of the \atlas observations were carried out under $\sim$1--2-arcsecond seeing conditions \citep{Emsellem_2004} and so the innermost radial points of each profile in \autoref{fig:ATLAS3D} may suffer accordingly.  Nevertheless, we see that the steep rise of the systemic velocity with radius, as well as the large difference in systemic velocities as measured at small and large radii, are both considerably more extreme in \j0437 than in the majority of \atlas galaxies. At about 200\,pc from the center the systemic velocity of a typical fast rotator deviates by only $\sim$2-3\,\kms from the systemic velocity measured near the center. This trend does not change substantially with increasing radius.

There are a few galaxies in the \atlas sample that have systemic velocity deviations similar to or even larger than those seen in \j0437, albeit at larger radii. The galaxies with the top four largest deviations are labeled in \autoref{fig:ATLAS3D}: NGC\,4753, UGC\,09519, NGC\,4342 and NGC\,3665.  NGC\,4753, which has the largest difference in the systemic velocity, also shows clear morphological evidence of a recent merger and contains complex dust filaments \citep{Krajnovic_2011,2020MNRAS.498.2138B}, indicating that it is likely not in equilibrium.  UGC\,09519 might be dusty in the center, and it also has an unusual large-scale stellar disk characterised by blue colours and low surface brightness \citep{2015MNRAS.446..120D}. NGC\,3665 has a well defined nuclear dust and gas disk \citep{2017MNRAS.468.4663O}, as well as asymmetric outer isophotes \citep{2020MNRAS.498.2138B}. NGC\,4342 shows no evidence for disturbances in morphology or kinematics, except harbouring a central nuclear stellar disk \citep{1998MNRAS.300..469S}, though \citet{1999ApJ...514..704C} note that this galaxy has a remarkably large central velocity dispersion for its size and luminosity. The \atlas sample is perhaps not an ideal comparison sample, as it is made of early-type galaxies and the observations do not probe the same spatial scales as the NIFS data of \j0437.  Nevertheless, it is clear that the majority of \atlas galaxies do not show strong variations in systemic velocity with radius, and there is some evidence that those with strong variations tend to exhibit other indications of kinematic disturbance.

\section{Observational constraints on the properties of a hypothetical binary SMBH system in \j0437} \label{app:Binary}

\begin{figure*}[t]
	\centering
		\includegraphics[width=0.60\textwidth]{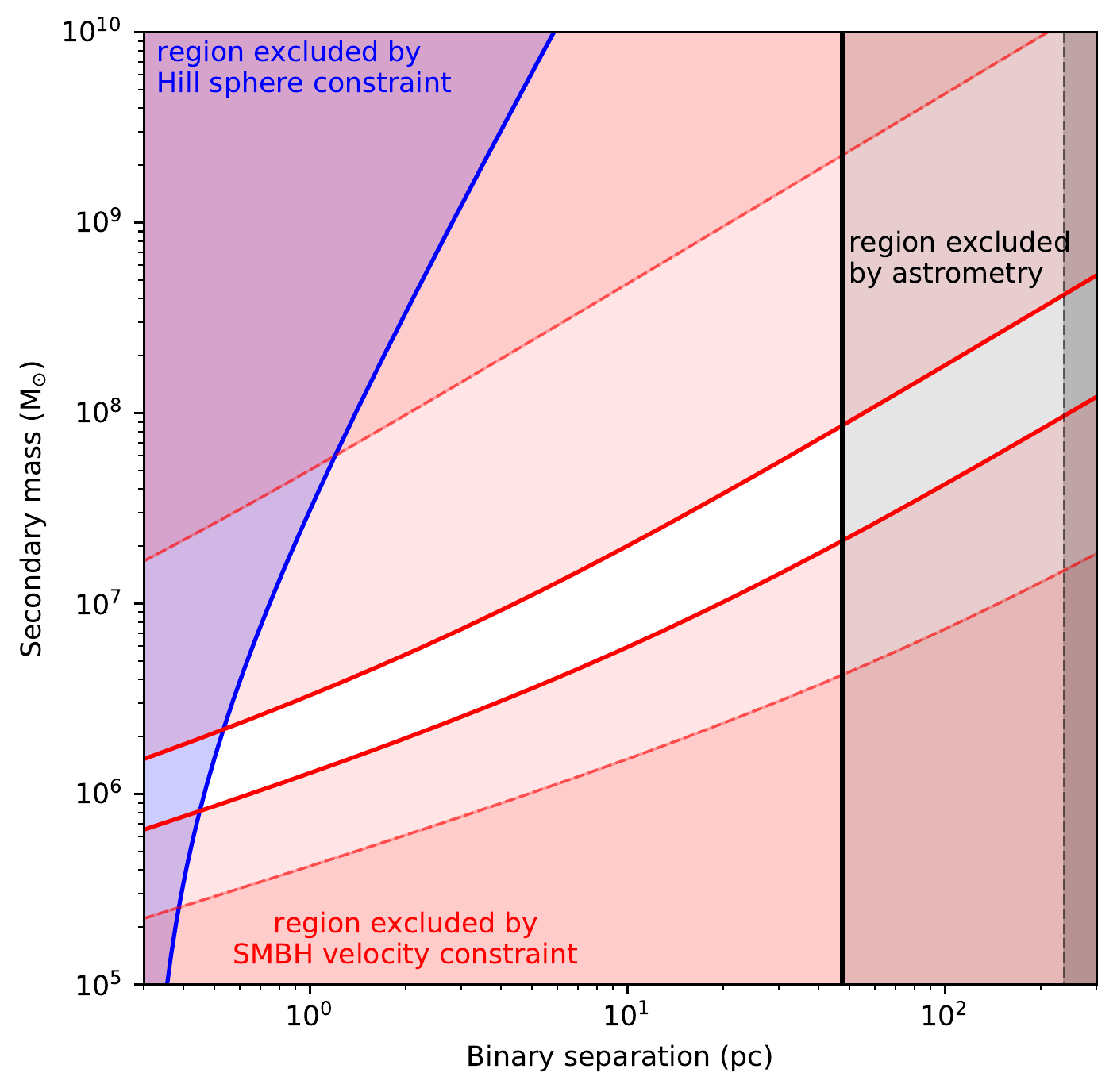}
	\caption{Observational constraints on the space of secondary SMBH mass and binary separation for \j0437.  The blue shaded region is excluded by the requirement that the observed maser disk be tidally undisrupted, the red shaded region is excluded by the requirement that the observed SMBH exhibit the measured velocity offset, and the gray shaded region is excluded by the lack of an astrometric offset seen between the SMBH and the galactic center.  The remaining unshaded region indicates the permitted range of secondary SMBH mass and binary separation in the presence of these constraints.  For the red and gray shaded regions, the solid and dashed lines represent 50\% and 90\% probability bounds, respectively, determined as described in \autoref{app:Binary}.}
	\label{fig:binary_constraints}
\end{figure*}

We observe the SMBH in \j0437 to have a velocity offset with respect to its host galaxy, as determined using various different systemic velocity tracers (see \autoref{sec:VelMeasurements}).  One possible explanation for this velocity offset is that the observed SMBH is part of a binary black hole system with a second, unseen SMBH.  In this case, we have several observational constraints on the properties that such a binary system must have; these constraints are illustrated in \autoref{fig:binary_constraints}.

Our first constraint comes from the fact that the observed SMBH in \j0437 is surrounded by an accretion disk, which is traced by H$_2$O maser emission to extend out to radii of $\sim$0.3\,pc \citepalias[][see also the right panel of \autoref{fig:maser_rotation_curve}]{Gao_2017}.  If a second SMBH is present outside of this accretion disk\footnote{A second SMBH located within the innermost observed edge of the accretion disk would likely go undetected by the maser measurements (such a tight binary system would appear to the maser system as a single SMBH with a mass equal to the combined masses of both SMBHs), but it would not by itself lead to an observed velocity offset between the maser measurements and the systemic velocity of the host galaxy.}, then its mass and separation from the observed SMBH must be such that it avoids tidally disrupting the accretion disk.  This condition is roughly equivalent to requiring that the outer edge of the accretion disk lie within the Hill sphere of the observed SMBH.  If we denote the mass of the observed SMBH as $m_1$, the mass of the second SMBH as $m_2$, their separation as $r$, and the Hill sphere radius as $r_H$, then we can cast this condition as an upper bound on $m_2$ of

\begin{equation}
m_2 \leq \frac{m_1}{r_H^2} \left[ \frac{1}{\left( r - r_H \right)^2} - \frac{1}{r^2} \right]^{-1} . \label{eqn:HillSphere}
\end{equation}

\noindent The blue shaded region in \autoref{fig:binary_constraints} shows the combinations of $m_2$ and $r$ that are excluded by this criterion.  We use the measured value of $m_1 = 2.9 \times 10^6$\,M$_{\odot}$ from \citetalias{Gao_2017} for the mass of the observed SMBH and the aforementioned value of $r_H = 0.3$\,pc from the VLBI map.

Our second constraint comes from the observed velocity offset of the SMBH with respect to the host galaxy.  If this SMBH is participating in a binary system, then its line-of-sight velocity $v$ is related to the parameters of the binary orbit via \citep[see, e.g.,][]{Murray_1999}

\begin{equation}
v = m_2 \sin(i) \big[ \cos(\omega + f) + e \cos(\omega) \big] \sqrt{\frac{2 G}{r \left( 1 - e^2 \right) \left( m_1 + m_2 \right)}} .
\end{equation}

\noindent Here, $i$ is the inclination of the orbital plane, $\omega$ is its argument of pericenter, $f$ is the true anomaly of the observed SMBH, and $e$ is the orbital eccentricity; $m_1$, $m_2$, and $r$ are the same as in \autoref{eqn:HillSphere}.  We do not currently have any ability to constrain the geometric parameters of the orbit, so we instead treat them probabilistically; we assume that the orbital plane is oriented randomly on the sphere (i.e., $\omega$ is distributed uniformly on $[0,2\pi]$ and $\cos(i)$ is distributed uniformly on $[-1,1]$), that $f$ is oriented randomly on the circle, and that $e$ is distributed uniformly in the range $[0,1]$.  The solid and dashed red lines in \autoref{fig:binary_constraints} show the 50\% and 90\% probability contours, respectively, for the combined constraints on $m_2$ and $r$ given these assumptions about the distribution of possible orbit geometries.  For the purposes of this constraint, we estimate the orbital velocity of the SMBH in \j0437 to be $48 \text{\,\kms} \leq v \leq 101$\,\kms based on the measurements presented in this paper (see \autoref{tab:Velocities}).

Our third and final constraint comes from the apparent lack of an astrometric offset between the SMBH in \j0437 and the center-of-light, determined in \citetalias{Pesce_2018} to be $\lesssim$0.05 arcseconds.  If we take this value to be an upper limit on the SMBH binary on-sky separation, then we can convert it into a constraint on the SMBH binary absolute separation.  We are again faced with the fact that we do not have any handle on the orientation of the binary orbit, so we assume that the orbit is randomly distributed on the sphere and plot 50\% and 90\% probability regions in \autoref{fig:binary_constraints} (shown as light and dark gray shaded regions, respectively).

\bibliography{bib}{}
\bibliographystyle{aasjournal}

\end{document}